\documentclass[traditabstract]{aa}
\usepackage{graphicx}
\usepackage{txfonts}
\usepackage{stfloats}     % to include figure spanning two columns at the bottom of a page
\newcommand{\minp}{\mbox{\rlap{.}$'$}} 
\newcommand{\tablenotea}[1]{\parbox{  8.9cm}{\indent \footnotesize{#1}}}
\newcommand{\tablenoteb}[1]{\parbox{18.4cm}{\indent \footnotesize{#1}}}
\newcommand{\tablerefs}[1]{\parbox{18.4cm}{ \indent \footnotesize{\textsc{References.--}~#1}}}

\newcommand{\jms}{J. Mol. Spectr.}
\newcommand{\jmst}{J. Mol. Struct.}

\newcommand{\cpl}{Chem. Phys. Lett.}

\newcommand{\chemrev}{Chem. Rev.}
\newcommand{\fdis}{Faraday Discussions}
\newcommand{\pecs}{Porc. Energy Combust. Sci.}
\newcommand{\jpcrd}{J. Phys. Chem. Ref. Data}
\newcommand{\energy}{Energy}
\newcommand{\bbpc}{Ber. Bunsenges. Phys. Chem.}

\begin{document}
\title{Discovery of interstellar ketenyl (HCCO),\\ a surprisingly abundant radical\thanks{Based on observations carried out with the IRAM 30m Telescope. IRAM is supported by INSU/CNRS (France), MPG (Germany) and IGN (Spain).}}
\titlerunning{Interstellar ketenyl}
\authorrunning{Ag\'undez et al.}

\author{
Marcelino Ag\'undez\inst{1},
Jos\'e Cernicharo\inst{1}, and
Michel Gu\' elin\inst{2}
}

\institute{
Instituto de Ciencia de Materiales de Madrid, CSIC, C/ Sor Juana In\'es de la Cruz 3, 28049 Cantoblanco, Spain \and
Institut de Radioastronomie Millim\'etrique, 300 rue de la Piscine, 38406 St. Martin d'H\'eres, France
}

\date{Received; accepted}

% \abstract{}{}{}{}{}
% 5 {} token are mandatory

\abstract
{We have conducted radioastronomical observations of 9 dark clouds with the IRAM 30m telescope. We present the first identification in space of the ketenyl radical (HCCO) toward the starless core Lupus-1A and the molecular cloud L483, and the detection of the related molecules ketene (H$_2$CCO) and acetaldehyde (CH$_3$CHO) in these two sources and 3 additional dark clouds. We also report the detection of the formyl radical (HCO) in the 9 targeted sources and of propylene (CH$_2$CHCH$_3$) in 4 of the observed sources, which extends significantly the number of dark clouds where these molecules are known to be present. We derive a beam-averaged column density of HCCO of $\sim 5 \times 10^{11}$ cm$^{-2}$ in both Lupus-1A and L483, which means that the ketenyl radical is just $\sim$10 times less abundant than ketene in these sources. The non-negligible abundance of HCCO found implies that there must be a powerful formation mechanism able to counterbalance the efficient destruction of this radical through reactions with neutral atoms. The column densities derived for HCO, (0.5-2.7) $\times 10^{12}$ cm$^{-2}$, and CH$_2$CHCH$_3$, (1.9-4-2) $\times 10^{13}$ cm$^{-2}$, are remarkably uniform across the sources where these species are detected, confirming their ubiquity in dark clouds. Gas phase chemical models of cold dark clouds can reproduce the observed abundances of HCO, but cannot explain the presence of HCCO in Lupus-1A and L483 and the high abundances derived for propylene. The chemistry of cold dark clouds needs to be revised in the light of these new observational results.}
{}
{}
{}
{}


\maketitle

\section{Introduction}

Organic molecules are ubiquitous in interstellar clouds. The most complex and saturated ones are found in clouds around young stellar objects, where they are most likely formed on the surface of dust grains and released to the gas phase by thermal evaporation (\cite{her2009} 2009). In cold dark clouds, the chemical composition is characterized by highly unsaturated carbon chains of the families of polyynes and cyanopolyynes and relatively simple oxygen-bearing organic molecules, whose synthesis relies to a large extent on gas phase chemical processes (\cite{agu2013} 2013). However, the widespread ocurrence of methanol (CH$_3$OH) in cold dark clouds, and the more recent detections of other complex and saturated organic molecules such as propylene (CH$_2$CHCH$_3$), methyl formate (CH$_3$OCOH), dimethyl ether (CH$_3$OCH$_3$), methoxy (CH$_3$O), and formic acid (HCOOH) in some dark clouds (\cite{mar2007} 2007; \cite{obe2010} 2010; \cite{bac2012} 2012; \cite{cer2012} 2012) have put on the table the role of grain surface reactions and non-thermal desorption processes in these cold and quiescent environments.

Many of the oxygen-bearing organic molecules observed in dark clouds can be described by the general formula H$_x$C$_n$O, with $n=1$ (CO, HCO, H$_2$CO, CH$_3$O, and CH$_3$OH), $n=2$ (C$_2$O, H$_2$CCO, and CH$_3$CHO), and $n=3$ (C$_3$O and HCCCHO). The formation of most of them is reasonably well explained by gas phase chemical reactions, with notable exceptions as in the case of CH$_3$OH (\cite{agu2013} 2013). A better understanding of the chemistry of dark clouds must necessarily entail deep observational studies able to enlarge both the number of sources chemically characterized and the inventory of molecules identified. The recent discovery of a starless core in the Lupus molecular cloud with a chemical richness comparable to that of the widely studied cloud TMC-1 (\cite{sak2009} 2009, 2010), provides an excellent opportunity to bring new observational constraints to the chemistry of dark clouds.

In this Letter we present the first detection in space of the ketenyl radical (HCCO), a missing link in the series H$_x$C$_2$O, in the starless core Lupus-1A and the molecular cloud L483. We also report observations of the more hydrogenated derivatives H$_2$CCO and CH$_3$CHO, the HCO radical, and the highly saturated hydrocarbon CH$_2$CHCH$_3$ in various molecular clouds.

\section{Observations}

The observations were carried out with the IRAM 30m telescope from September to November 2014 during an observational campaign aimed at searching for negative ions in molecular sources. The observed sources are Lupus-1A (\cite{sak2010} 2010), L483 (\cite{agu2008} 2008), L1521F (\cite{cra2005} 2005), the Serpens South complex at the HC$_7$N emission peak 1a (\cite{fri2013} 2013), L1495B, L1389, L1172, L1251A, and L1512 (\cite{cor2013} 2013). The coordinates of the sources were taken from the references quoted above. We used the EMIR receivers in single sideband mode with image rejections $>$10 dB to cover selected frequency ranges from 83 to 105 GHz (HPBW in the range 23-29$''$). System temperatures ranged from 80 to 150 K. The highest values were obtained for Lupus-1A due to the low elevation ($<$20$^{\circ}$) reached by this source at the IRAM 30m latitude. We used the FFTS backend in its narrow mode, providing a bandwidth of 1.8 GHz and a spectral resolution of 50 kHz, which translates to velocity resolutions of 0.14-0.18 km\,s$^{-1}$ at the observed frequencies. We used the frequency switching technique with a frequency throw of 7.8 MHz to minimize the effects of the standing waves between the secondary mirror and the receivers on the spectral baselines. Pointing and focus were regularly checked every 1-2 h by observing nearby planets or quasars.

\section{Results}

\begin{figure}
\centering
\includegraphics[angle=0,width=\columnwidth]{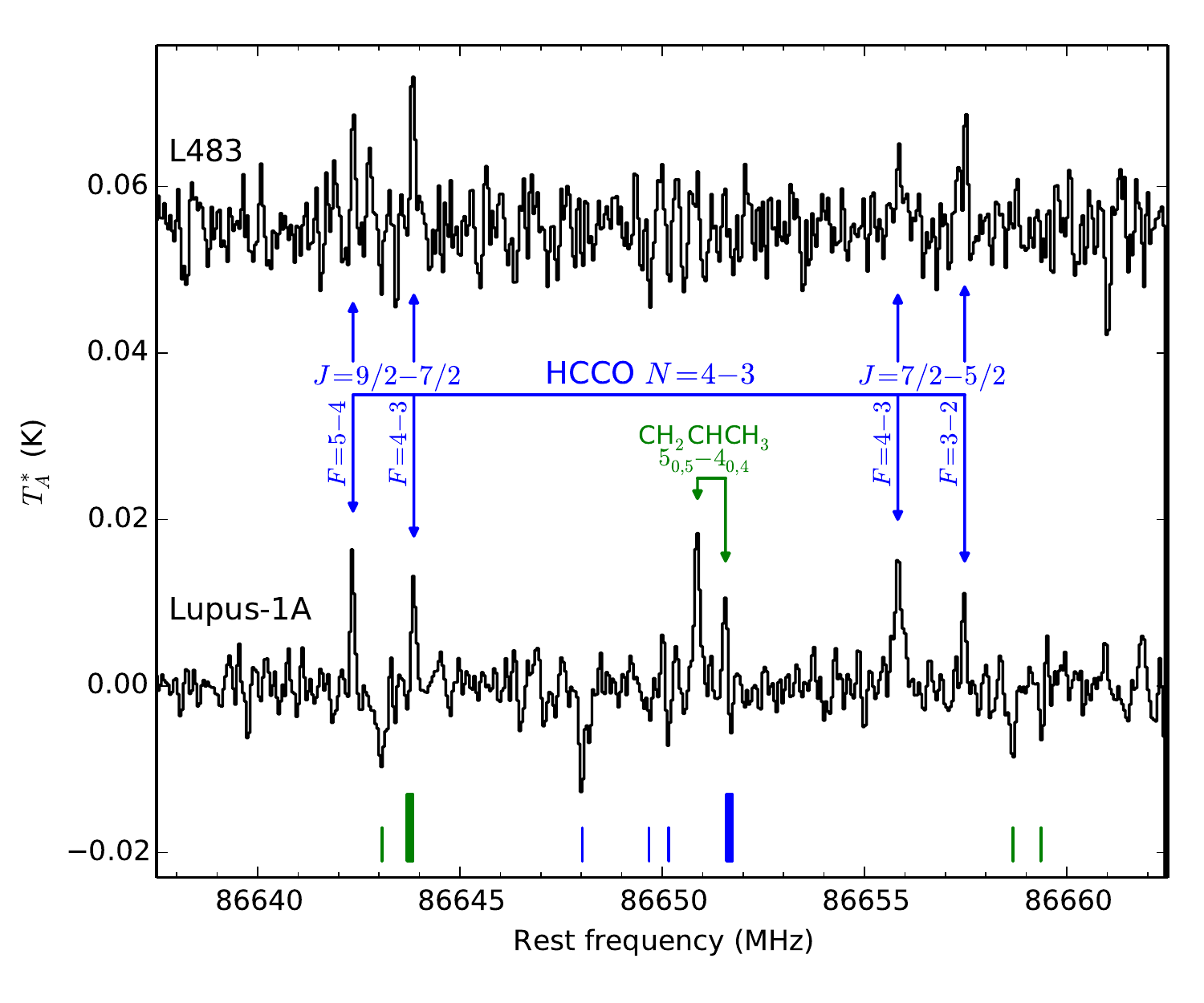}
\caption{Spectra of Lupus-1A and L483 around 86.65 GHz showing the quartet of emission lines assigned to HCCO and the doublet assigned to CH$_2$CHCH$_3$ in Lupus-1A. The rest frequency scale corresponds to a systemic velocity $V_{\rm LSR}$ = +5.0 km\,s$^{-1}$ (Lupus-1A) and +5.3 km\,s$^{-1}$ (L483). The $T_A^*$ rms is 0.0022 K (Lupus-1A) and 0.0034 K (L483) per 50 kHz channel. The bottom vertical marks indicate the position of the frequency switching negative artifacts, located at $\pm$7.8 MHz from each line. Note that in the Lupus-1A spectrum, the intensity of the HCCO $J=9/2-7/2$ $F=4-3$ and the A CH$_2$CHCH$_3$  $5_{0,5}-4_{0,4}$ lines, whose frequencies differ by 7.7 MHz, is artificially reduced due to an overlap with frequency switching negative artifacts (thick bottom marks).} \label{fig:lines}
\end{figure}

The spectrum of Lupus-1A around 86.65 GHz shows a quartet of emission lines whose frequencies coincide precisely with the strongest fine and hyperfine components of the $N=4-3$ rotational transition of the HCCO radical (see Fig.~\ref{fig:lines} and Table~\ref{table:lines}). The lines, with antenna temperatures of 0.010-0.015 K, are detected at $\sim$10$\sigma$ confidence after 7 h of `on source' telescope time. The same quartet of lines is detected in L483 at a lower confidence level of $\sim$5$\sigma$ (see Fig.~\ref{fig:lines} and Table~\ref{table:lines}).

The ketenyl radical has a planar structure, with a practically linear CCO backbone and the H atom lying out of the linear axis. The radical has a $^2A''$ ground electronic state and a complex rotational structure whose $N_{K_a,K_c}$ levels split in a fine (electronic spin-rotation interaction) and hyperfine (H nuclear spin) structure described by the quantum numbers $J$ and $F$, respectively (\cite{end1987} 1987; \cite{ohs1993} 1993; \cite{sat2004} 2004). We obtained the line frequencies from the Cologne Database for Molecular Spectroscopy\footnote{See \texttt{http://www.astro.uni-koeln.de/cdms/}} (\cite{mul2005} 2005), where only the levels $K_a=0$ are considered, in which case the effective hamiltonian is similar to that of a linear molecule with a $^2\Sigma$ state. The total dipole moment, which should be only marginally larger than the component along the $a$ axis, has been calculated as 1.59 D by \cite{sza1996} (1996) and as 1.68 D by \cite{jer2007} (2007). We adopt this latter value for $\mu_a$.

\begin{table}
\caption{Observed line parameters of the $N=4-3$ transition of HCCO} \label{table:lines}
\centering
\begin{tabular}{lrcll}
\hline \hline
\multicolumn{1}{c}{Transition} & \multicolumn{1}{c}{Frequency} & \multicolumn{1}{c}{$V_{\rm LSR}$} & \multicolumn{1}{c}{$\Delta v$}      & \multicolumn{1}{c}{$\int T_A^* dv$} \\
\multicolumn{1}{c}{$J'~~F' - J''~~F''$} & \multicolumn{1}{c}{(MHz)}                 & \multicolumn{1}{c}{(km\,s$^{-1}$)}    & \multicolumn{1}{c}{(km\,s$^{-1}$)} & \multicolumn{1}{c}{(K\,km\,s$^{-1}$)} \\
\hline
\multicolumn{5}{c}{Lupus-1A} \\
\hline
9/2~~5 $-$ 7/2~~4 & 86642.357 & +5.05(4) & 0.36(8)  & 0.006(1) \\
9/2~~4 $-$ 7/2~~3 & 86643.862 & +5.01(5) & 0.42(10)& 0.006(1)$^a$ \\
7/2~~4 $-$ 5/2~~3 & 86655.825 & +4.96(5) & 0.71(14)& 0.011(2) \\
7/2~~3 $-$ 5/2~~2 & 86657.477 & +5.05(5) & 0.35(10)& 0.004(1) \\
\hline
\multicolumn{5}{c}{L483} \\
\hline
9/2~~5 $-$ 7/2~~4 & 86642.357 & +5.27(4) & 0.35(9)  & 0.006(1) \\
9/2~~4 $-$ 7/2~~3 & 86643.862 & +5.39(4) & 0.42(8)  & 0.009(1) \\
7/2~~4 $-$ 5/2~~3 & 86655.825 & +5.20(5) & 0.47(15) & 0.005(1) \\
7/2~~3 $-$ 5/2~~2 & 86657.477 & +5.22(4) & 0.36(11) & 0.006(1) \\
\hline
\end{tabular}
\tablenotea{\\
Numbers in parentheses are 1$\sigma$ uncertainties in units of the last digits. Line parameters obtained by gaussian fits to the line profiles.\\
$^a$ Intensity is reduced due to overlap with frequency switching artifact.
}
\end{table}

\begin{table*}
\caption{Column densities (in cm$^{-2}$) and column density ratios} \label{table:columndensities}
\centering
\tiny
%\scriptsize
\begin{tabular}{lrrcrcrrrr}
\hline \hline
& & & \multicolumn{1}{c}{\scriptsize{$N$(HCCO)}} & & \multicolumn{1}{c}{\scriptsize{$N$(H$_2$CCO)}} & & & \multicolumn{1}{c}{\scriptsize{$N$(HCO)}} & \\
\cline{4-4} \cline{6-6} \cline{9-9}
\multicolumn{1}{l}{Source} & \multicolumn{1}{r}{\scriptsize{$N$(H$_2$)~~~~~\tiny{Ref}}} & \multicolumn{1}{c}{\scriptsize{$N$(HCCO)}} &\multicolumn{1}{c}{\scriptsize{$N$(H$_2$)}} & \multicolumn{1}{c}{\scriptsize{$N$(H$_2$CCO)}} & \multicolumn{1}{c}{\scriptsize{$N$(HCCO)}} & \multicolumn{1}{c}{\scriptsize{$N$(CH$_3$CHO)}} & \multicolumn{1}{c}{\scriptsize{$N$(HCO)}} & \multicolumn{1}{c}{\scriptsize{$N$(H$_2$)}} & \multicolumn{1}{c}{\scriptsize{$N$(CH$_2$CHCH$_3$)}} \\
\hline
Lupus-1A               & $1.5 \times 10^{22}$~~(1)   & $5.4 \times 10^{11}$     & $3.6 \times 10^{-11}$   & $4.3 \times 10^{12}$ & 8                                & $2.3 \times 10^{12}$  & $2.5 \times 10^{12}$ & $1.7 \times 10^{-10}$ & $2.6 \times 10^{13}$ \\
L483                      & $3.0 \times 10^{22}$~~(2)   & $4.3 \times 10^{11}$     & $1.4 \times 10^{-11}$   & $4.4 \times 10^{12}$ & 10                              & $3.2 \times 10^{12}$  & $2.7 \times 10^{12}$ & $9.0 \times 10^{-11}$ & $< 1.0 \times 10^{13}$ \\
L1495B                  & $1.2 \times 10^{22}$~~(3)   & $< 1.3 \times 10^{11}$ & $< 1.1 \times 10^{-11}$ & $2.2 \times 10^{12}$ & $>17$                         & $5.8 \times 10^{11}$  & $6.7 \times 10^{11}$ & $5.6 \times 10^{-11}$ & $2.8 \times 10^{13}$ \\
L1521F                  & $1.35 \times 10^{23}$~~(4) & $< 1.0 \times 10^{11}$ & $< 7.4 \times 10^{-13}$ & $2.8 \times 10^{12}$ & $>28$                         & $7.9 \times 10^{11}$  & $1.7 \times 10^{12}$ & $1.3 \times 10^{-11}$ & $1.9 \times 10^{13}$ \\
Serpens South 1a & $2.0 \times 10^{22}$~~(5)   & $< 1.7 \times 10^{11}$ & $< 8.5 \times 10^{-12}$ & $3.9 \times 10^{12}$  & $>23$                         & $1.0 \times 10^{12}$ & $1.6 \times 10^{12}$ & $8.0 \times 10^{-11}$ & $4.2 \times 10^{13}$ \\
L1389                    & $1.0 \times 10^{23}$~~(6)   & \multicolumn{1}{c}{--}   & \multicolumn{1}{c}{--}    & \multicolumn{1}{c}{--} & \multicolumn{1}{c}{--} & \multicolumn{1}{c}{--} & $2.0 \times 10^{12}$ & $2.0 \times 10^{-11}$ & \multicolumn{1}{c}{--} \\
L1172                    & $6.0 \times 10^{21}$~~(7)   & \multicolumn{1}{c}{--}   & \multicolumn{1}{c}{--}    & \multicolumn{1}{c}{--} & \multicolumn{1}{c}{--} & \multicolumn{1}{c}{--} & $6.3 \times 10^{11}$ & $1.1 \times 10^{-10}$ & \multicolumn{1}{c}{--} \\
L1251A                  & $5.1 \times 10^{21}$~~(8)   & \multicolumn{1}{c}{--}   & \multicolumn{1}{c}{--}    & \multicolumn{1}{c}{--} & \multicolumn{1}{c}{--} & \multicolumn{1}{c}{--} & $6.0 \times 10^{11}$ & $1.2 \times 10^{-10}$ & \multicolumn{1}{c}{--} \\
L1512                    & $5.7 \times 10^{21}$~~(3)   & \multicolumn{1}{c}{--}   & \multicolumn{1}{c}{--}    & \multicolumn{1}{c}{--} & \multicolumn{1}{c}{--} & \multicolumn{1}{c}{--} & $4.6 \times 10^{11}$ & $8.1 \times 10^{-11}$ & \multicolumn{1}{c}{--} \\
\hline
\end{tabular}
\tablenoteb{\\
Upper limits are given at 3$\sigma$ confidence. A blank value indicates that observations were not sensitive enough.
}
\tablerefs{
(1) This study; (2) \cite{taf2000} (2000); (3) \cite{mye1983} (1983); (4) \cite{cra2005} (2005); (5) \cite{fri2013} (2013); (6) \cite{lau2010} (2010); (7) quoted by \cite{cor2013} (2013); (8) \cite{sat1994} (1994).
}
\end{table*}

In Lupus-1A, we derive a kinetic temperature of 14 $\pm$ 2 K from observations of CH$_3$CCH $J=5-4$ ($K=0,1,2$). At the position of IRAS15398$-$3359, 2\minp8 away from Lupus-1A, \cite{sak2009} (2009) obtain 12.6 $\pm$ 1.5 K. In the Lupus-1A core, \cite{sak2010} (2010) derive a rotational temperature of 7.3 $\pm$ 1 K for C$_6$H, while we derive 8.2 $\pm$ 0.8 K and 5.5 $\pm$ 0.8 K for C$_4$H and C$_3$N, respectively, from observations of a couple of doublets of lines. Since the dipole moment of HCCO is in between those of C$_4$H and C$_3$N we expect a rotational temperature in between 5.5 and 8.2 K. Adopting a rotational temperature of 7 K, we find a beam-averaged column density of $5.4 \times 10^{11}$ cm$^{-2}$ for HCCO (only $K_a=0$ levels) in Lupus-1A. In the dense core L483, adopting a rotational temperature of 10 K, close to the gas kinetic temperature (\cite{ful1993} 1993; \cite{ang1997} 1997), we derive a beam-averaged column density of $4.3 \times 10^{11}$ cm$^{-2}$. From a previous unsuccessful search for HCCO in interstellar clouds, \cite{tur1989} (1989) reported an upper limit to its column density of $2.9 \times 10^{10}$ cm$^{-2}$ in TMC-1, although from their observational details we derive an upper limit of $5 \times 10^{11}$ cm$^{-2}$ in TMC-1, i.e., close to the values derived in Lupus-1A and L483. The difference may arise from the different dipole moment and spectroscopic parameters adopted by these authors and us. In the sources in which we obtained sensitive observations around 86.65 GHz (L1495B, L1521F, and Serpens South 1a) we derive 3$\sigma$ upper limits to the column density of HCCO around 10$^{11}$ cm$^{-2}$, for a rotational temperature of 10 K (see Table~\ref{table:columndensities}).

The column densities of H$_2$ in the observed sources, which are needed to convert molecular column densities into fractional abundances relative to H$_2$, are compiled in Table~\ref{table:columndensities}. \cite{sak2009} (2009) estimate a column density of H$_2$ of $3 \times 10^{22}$ cm$^{-2}$ toward IRAS15398$-$3359, located 2\minp8 away from Lupus-1A, based on the similar dust continuum flux at 850 and 450 $\mu$m of the IRAS source and L1527. In our observations we have detected the $J=1-0$ line of $^{13}$C$^{18}$O in Lupus-1A, with a velocity-integrated antenna temperature of $0.037 \pm 0.006$ K km s$^{-1}$, which translates to a column density of $5.7 \times 10^{13}$ cm$^{-2}$ assuming a rotational temperature of 10 K. Adopting the $^{16}$O/$^{18}$O ratio of 500 within the local interstellar medium (\cite{zin1996} 1996), the relationship between $N$($^{13}$CO) and $A_V$ recommended by \cite{tot2009} (2009) for the Lupus molecular cloud, and the canonical relationship between $A_V$ and $N$(H$_2)$ of \cite{boh1978} (1978), we derive $N$(H$_2$) = $1.5 \times 10^{22}$ cm$^{-2}$ in Lupus-1A. This value could be somewhat higher if the densities are high enough to result in a significant depletion of CO. In the dense core L483, \cite{ang1997} (1997) derive $N$(H$_2$) = $1.4 \times 10^{22}$ cm$^{-2}$ from NH$_3$ observations, while \cite{taf2000} (2000) derive $N$(H$_2$) = $3 \times 10^{22}$ cm$^{-2}$ from C$^{17}$O observations. The much higher value of $9.3 \times 10^{23}$ cm$^{-2}$ derived by \cite{jor2002} (2002) from dust continuum flux measurements may correspond to a more compact region of high column density than probed by molecular observations. The fractional abundance relative to H$_2$ of the HCCO radical is then a few 10$^{-11}$ in both Lupus-1A and L483.

We have observed in Lupus-1A, L483, L1495B, L1521F, and Serpens South 1a, a couple of additional members of the series H$_x$C$_2$O: ketene (H$_2$CCO) and the more saturated species acetaldehyde (CH$_3$CHO), two molecules that have been previously found in various cold dark clouds (\cite{mat1985} 1985; \cite{irv1989} 1989; \cite{ohi1998} 1998; \cite{obe2010} 2010; \cite{bac2012} 2012; \cite{cer2012} 2012). The line parameters are listed in Tables~\ref{table:h2cco} and \ref{table:ch3cho} and the column densities are given in Table~\ref{table:columndensities}. The column densities of ketene and acetaldehyde in the five sources where they have been detected are quite uniform, in the range (0.6-4.4) $\times 10^{12}$ cm$^{-2}$, and similar to those found in TMC-1 (see, e.g., Table~4 of \cite{agu2013} 2013). We find that the ketenyl radical is 8-10 times less abundant than ketene in Lupus-1A and L483, while in L1495B, L1521F, and Serpens South 1a, as well as in TMC-1, HCCO is more than 10 times less abundant than H$_2$CCO.

The formyl radical (HCO) has been detected in the 9 targeted sources through the 4 strongest components of the $1_{0,1}-0_{0,0}$ transition, lying at 86 GHz. In 4 of the sources, only one component was detected because the other three were not covered by the observations (see Table~\ref{table:hco}). The HCO lines are relatively intense in all observed sources, with antenna temperatures in the range 0.03-0.2 K. The presence of the formyl radical in dark clouds was first reported by \cite{jim2004} (2004) in L1448 and later on by \cite{cer2012} (2012) in B1 and TMC-1. We derive column densities of HCO which are remarkably uniform across the observed sources, with values in the range (0.5-2.7) $\times 10^{12}$ cm$^{-2}$ (see Table~\ref{table:columndensities}), similar to the value derived in B1 (\cite{cer2012} 2012) and slightly above that found in L1448 (\cite{jim2004} 2004). It is also worth noting that the fractional abundances of HCO relative to H$_2$ are quite uniform, in the range 10$^{-11}$-10$^{-10}$, across all sources (see Table~\ref{table:columndensities}).

In the same spectrum of Lupus-1A that shows the quartet of lines of HCCO (see Fig.~\ref{fig:lines}), there is a doublet of lines arising from propylene (CH$_2$CHCH$_3$), a highly saturated hydrocarbon which has been previously found only in TMC-1 (\cite{mar2007} 2007). Propylene has been detected in Lupus-1A and three additional dense clouds: L1495B, L1521F, and Serpens South 1a (see Table~\ref{table:ch2chch3}). In some of the sources, only the strongest transitions were clearly detected above the noise level. As in TMC-1, the column densities derived in the 4 sources are a few 10$^{13}$ cm$^{-2}$ (see Table~\ref{table:columndensities}) and are consistent with a A/E ratio of 1. The fractional abundances relative to H$_2$ are in the range (0.14-2.3) $\times 10^{-9}$.

\section{Discussion}

After the astronomical search by \cite{tur1989} (1989), the ketenyl radical has not attracted much attention in astrochemistry, although it has been of some interest in combustion chemistry (\cite{bau1992} 1992) and in the study of Titan's atmosphere (\cite{dob2014} 2014). No reaction involving this species is included in any of the main chemical kinetics databases of use in interstellar chemistry, UMIST (\cite{mce2013} 2013) and KIDA (\cite{wak2015} 2015). The detection of HCCO in Lupus-1A makes it worth discussing its chemistry in interstellar clouds.

\begin{figure}
\centering
\includegraphics[angle=0,width=\columnwidth]{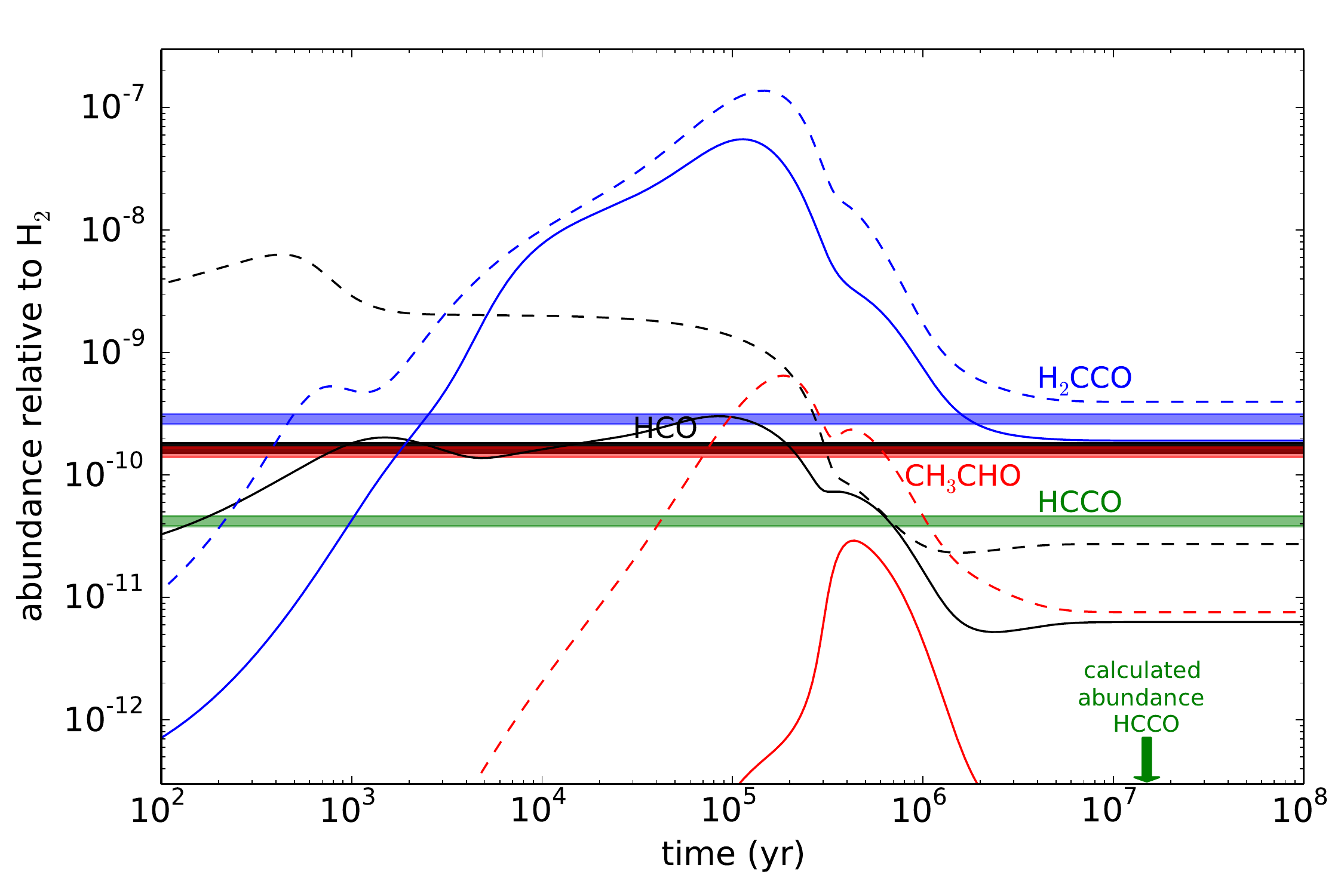}
\caption{Calculated abundances as a function of time for a standard dark cloud model (parameters from \cite{agu2013} 2013). Solid lines use the KIDA \texttt{kida.uva.2014} ratefile (\cite{wak2015} 2015) and dashed lines the UMIST \texttt{RATE12} ratefile (\cite{mce2013} 2013). In both cases a subset of reactions involving HCCO is introduced, although its calculated abundance remains below the range plotted. The horizontal thick lines indicate the observed abundances in Lupus-1A.} \label{fig:abun}
\end{figure}

By including various formation and destruction reactions of HCCO from combustion chemistry and the literature in chemical kinetics in a standard chemical model of a cold dark cloud (see, e.g., \cite{agu2013} 2013) we find that the main formation route to HCCO is the reaction between OH and C$_2$H. This reaction is exothermic and has an estimated rate constant of $3 \times 10^{-11}$ cm$^{3}$\,s$^{-1}$ (\cite{fre1992} 1992). The reaction between atomic O and C$_2$H$_2$ is also exothermic in the channel yielding HCCO, although its reaction barrier is too large to be relevant at low temperatures (\cite{bal2012} 2012). A possible precursor of HCCO is ketene. The reaction of ketene with OH is a potential source of HCCO radicals as this reaction is rapid (\cite{bro1989} 1989), although it seems to have a low yield of HCCO (\cite{gru1994} 1994). The ketenyl radical does not react with stable hydrocarbons or H$_2$ but it does react fast with radicals and atoms (\cite{dob2014} 2014). In dark clouds, the destruction of HCCO is dominated by the reactions with H, O, and N atoms, whose rate constants are of the order of 10$^{-10}$ cm$^{3}$\,s$^{-1}$ (\cite{bau1992} 1992; \cite{dob2014} 2014). The abundance of HCCO calculated by the chemical model remains about six orders of magnitude below that calculated for H$_2$CCO (see also \cite{occ2013} 2013), i.e., well below the values observed in Lupus-1A and L483. It is clear that some key formation mechanism is missing. In the chemical model, ketene, which is predicted with an abundance much larger than observed at an early time of $\sim$10$^5$ yr and in good agreement with the observed value at late times (see Fig.~\ref{fig:abun}), is mainly formed by the reaction of atomic O and the C$_2$H$_3$ radical and the dissociative recombination of the CH$_3$CO$^+$ ion with electrons. If we assume that these routes contribute also to the formation of HCCO, then its abundance increases significantly, but still remains 3-4 orders of magnitude below that of H$_2$CCO. 

A powerful formation mechanism of HCCO is needed to counterbalance the efficient depletion of this radical by reactions with neutral atoms. An interesting possibility are the reactions of O atoms with C$_n$H radicals ($n \geq 3$), which are assumed to yield CO as this is the most exothermic channel (\cite{smi2004} 2004), although the channel leading to HCCO is also exothermic. It is also worth noting that recent studies (\cite{hud2013} 2013; \cite{mai2014} 2014) find that ketene can be efficiently formed in various types of ices upon irradiation with energetic electrons and ultraviolet photons (both of which can be formed in dark clouds following cosmic-ray impacts), which opens the possibility to the formation of HCCO as an intermediate.

In the chemical model, the HCO radical is mainly formed through the reaction of O atoms and CH$_2$ radicals, together with the dissociative recombination of H$_3$CO$^+$ and H$_2$CO$^+$ with electrons, reaching an abundance which is in good agreement with the observed values (see Fig.~\ref{fig:abun}). On the other hand, the detection of propylene in various dark clouds other than TMC-1 points to a widespread presence of this saturated hydrocarbon in cold dark clouds. This fact exacerbates the challenge to chemical models, which still lack an efficient gas phase formation mechanism for this molecule (\cite{lin2013} 2013).

\section{Summary}

We have reported the first detection in space of the ketenyl radical (HCCO) in the starless core Lupus-1A and the molecular cloud L483, together with observations of the related molecules H$_2$CCO and CH$_3$CHO toward these two sources and some other dense clouds. Our observations also extend significantly the number of dark clouds in which the formyl radical (HCO), previously detected only in L1448, B1, and TMC-1, and propylene (CH$_2$CHCH$_3$), previously identified only in TMC-1, are observed. The radical HCCO is found to be just $\sim 10$ times less abundant than its closed shell counterpart H$_2$CCO, more than 3 orders of magnitude more abundant than predicted by chemical models. A highly efficient formation mechanism of HCCO, yet unidentified, is needed to counterbalance the rapid depletion of this radical in cold dark clouds by reactions with neutral atoms. The discovery of ketenyl in Lupus-1A and L483 and the widespread ocurrence of propylene in dark clouds suggest that a revision of the gas phase and grain surface chemical processes at work in cold dark clouds is required.

\begin{acknowledgements}

We thank Bel\'en Tercero for a critical reading of the manuscript, the anonymous referee for useful suggestions, and the IRAM 30m staff for their help during the observations. M.A. and J.C. thank funding support from the European Research Council (ERC Grant 610256: NANOCOSMOS) and from Spanish MINECO through grants CSD2009-00038, AYA2009-07304, and AYA2012-32032.

\end{acknowledgements}

\onltab{3}{
\begin{table*}
\caption{Observed line parameters of H$_2$CCO} \label{table:h2cco}
\flushleft
\tiny
\begin{tabular}{l@{\hspace{1.1cm}} | @{\hspace{0.1cm}}c@{\hspace{0.25cm}}c@{\hspace{0.25cm}}c | @{\hspace{0.1cm}}c@{\hspace{0.25cm}}c@{\hspace{0.25cm}}c | @{\hspace{0.1cm}}c@{\hspace{0.25cm}}c@{\hspace{0.25cm}}c@{\hspace{0.1cm}}}
\hline \hline
\multicolumn{1}{l}{Transition} & \multicolumn{3}{c}{$J_{K_a,K_c} = 5_{1,5} - 4_{1,4}$} & \multicolumn{3}{c}{$J_{K_a,K_c} = 5_{0,5} - 4_{0,4}$}  & \multicolumn{3}{c}{$J_{K_a,K_c} = 5_{1,4} - 4_{1,3}$} \\
\multicolumn{1}{l}{Frequency (MHz)} & \multicolumn{3}{c}{100094.514}                & \multicolumn{3}{c}{101036.630}                & \multicolumn{3}{c}{101981.429} \\
\hline
             & $V_{\rm LSR}$ & $\Delta v$ & $\int T_A^* dv$ & $V_{\rm LSR}$ & $\Delta v$ & $\int T_A^* dv$ & $V_{\rm LSR}$ & $\Delta v$ & $\int T_A^* dv$ \\
Source & (km/s) & (km/s) & (K km/s) & (km/s) & (km/s) & (K km/s) & (km/s) & (km/s) & (K km/s) \\
\hline
Lupus-1A               & +5.06(2) & 0.32(2) & 0.043(2)  & +5.06(3) & 0.41(7) & 0.036(5)  & +5.04(3) & 0.27(4) & 0.037(5) \\
L483                      & +5.28(1) & 0.46(2) & 0.069(2)  & -- & -- & --  & -- & -- & -- \\
L1495B                  & +7.62(1) & 0.30(1) & 0.035(1)  & -- & -- & --  & -- & -- & -- \\
L1521F                  & +6.39(1) & 0.39(1) & 0.044(1)  & -- & -- & --  & -- & -- & -- \\
Serpens South 1a & +7.58(1) & 0.64(2) & 0.062(2)  & -- & -- & --  & -- & -- & -- \\
\hline
\end{tabular}
\tablenoteb{\\
Numbers in parentheses are 1$\sigma$ uncertainties in units of the last digits. Line parameters obtained by gaussian fits to the line profiles.
}
\end{table*}
}% End onltab

\onltab{4}{
\begin{table*}
\caption{Observed line parameters of CH$_3$CHO} \label{table:ch3cho}
\centering
\tiny
\begin{tabular}{l@{\hspace{1.1cm}} | @{\hspace{0.1cm}}c@{\hspace{0.25cm}}c@{\hspace{0.25cm}}c | @{\hspace{0.1cm}}c@{\hspace{0.25cm}}c@{\hspace{0.25cm}}c | @{\hspace{0.1cm}}c@{\hspace{0.25cm}}c@{\hspace{0.25cm}}c@{\hspace{0.1cm}} | @{\hspace{0.1cm}}c@{\hspace{0.25cm}}c@{\hspace{0.25cm}}c}
\hline \hline
\multicolumn{1}{l}{Transition} & \multicolumn{3}{c}{$J_{K_a,K_c} = 2_{1,2} - 1_{0,1}$ E} & \multicolumn{3}{c}{$J_{K_a,K_c} = 2_{1,2} - 1_{0,1}$ A}  & \multicolumn{3}{c}{$J_{K_a,K_c} = 5_{1,4} - 4_{1,3}$ E}  & \multicolumn{3}{c}{$J_{K_a,K_c} = 5_{1,4} - 4_{1,3}$ A }     \\
\multicolumn{1}{l}{Frequency (MHz)} & \multicolumn{3}{c}{83584.279}                & \multicolumn{3}{c}{84219.748}                & \multicolumn{3}{c}{98863.313}                 & \multicolumn{3}{c}{98900.944}     \\
\hline
             & $V_{\rm LSR}$ & $\Delta v$ & $\int T_A^* dv$ & $V_{\rm LSR}$ & $\Delta v$ & $\int T_A^* dv$ & $V_{\rm LSR}$ & $\Delta v$ & $\int T_A^* dv$ & $V_{\rm LSR}$ & $\Delta v$ & $\int T_A^* dv$ \\
Source & (km/s) & (km/s) & (K km/s) & (km/s) & (km/s) & (K km/s) & (km/s) & (km/s) & (K km/s) & (km/s) & (km/s) & (K km/s) \\
\hline
Lupus-1A               & +5.02(9) & 0.76(19) & 0.008(2) & +4.98(3) & 0.30(6)   & 0.007(1) & +5.04(2) & 0.33(4)   & 0.009(1) & +5.03(2) & 0.30(4)   & 0.008(1) \\
L483                      & -- & -- & -- & +5.15(6) & 0.53(12)   & 0.012(2) & +5.27(2) & 0.78(7)   & 0.037(2) & +5.30(2) & 0.54(6)   & 0.027(2) \\
L1495B                  & -- & -- & -- & -- & -- & -- & +7.57(6) & 0.56(14)   & 0.006(1) & +7.40(10) & 0.77(19)   & 0.005(1) \\
L1521F                  & -- & -- & -- & -- & -- & -- & +6.46(4) & 0.51(11)   & 0.007(1) & +6.47(4) & 0.47(9)   & 0.008(1) \\
Serpens South 1a & -- & -- & -- & -- & -- & -- & +7.41(6) & 0.62(11)   & 0.009(2) & +7.53(6) & 0.66(15)   & 0.010(2) \\
\hline
\end{tabular}
\tablenoteb{\\
Numbers in parentheses are 1$\sigma$ uncertainties in units of the last digits. Line parameters obtained by gaussian fits to the line profiles.
}
\end{table*}
}

\onltab{5}{
\begin{table*}
\caption{Observed line parameters of HCO $N_{K_a,K_c} = 1_{0,1}-0_{0,0}$} \label{table:hco}
\centering
\tiny
\begin{tabular}{l@{\hspace{1.1cm}} | @{\hspace{0.1cm}}c@{\hspace{0.25cm}}c@{\hspace{0.25cm}}c | @{\hspace{0.1cm}}c@{\hspace{0.25cm}}c@{\hspace{0.25cm}}c | @{\hspace{0.1cm}}c@{\hspace{0.25cm}}c@{\hspace{0.25cm}}c@{\hspace{0.1cm}} | @{\hspace{0.1cm}}c@{\hspace{0.25cm}}c@{\hspace{0.25cm}}c}
\hline \hline
\multicolumn{1}{l}{Transition} & \multicolumn{3}{c}{$J=3/2-1/2$~~~~~$F=2-1$} & \multicolumn{3}{c}{$J=3/2-1/2$~~~~~$F=1-0$}  & \multicolumn{3}{c}{$J=1/2-1/2$~~~~~$F=1-1$}  & \multicolumn{3}{c}{$J=1/2-1/2$~~~~~$F=0-1$}     \\
\multicolumn{1}{l}{Frequency (MHz)} & \multicolumn{3}{c}{86670.76}                & \multicolumn{3}{c}{86708.36}                & \multicolumn{3}{c}{86777.46}                 & \multicolumn{3}{c}{86805.78}     \\
\hline
             & $V_{\rm LSR}$ & $\Delta v$ & $\int T_A^* dv$ & $V_{\rm LSR}$ & $\Delta v$ & $\int T_A^* dv$ & $V_{\rm LSR}$ & $\Delta v$ & $\int T_A^* dv$ & $V_{\rm LSR}$ & $\Delta v$ & $\int T_A^* dv$ \\
Source & (km/s) & (km/s) & (K km/s) & (km/s) & (km/s) & (K km/s) & (km/s) & (km/s) & (K km/s) & (km/s) & (km/s) & (K km/s) \\
\hline
Lupus-1A               & +5.23(1) & 0.38(2)  & 0.069(1)    & +5.34(1) & 0.44(2)  & 0.052(1)$^a$    & +5.22(1) & 0.38(2)  & 0.047(1)   & +5.27(3) & 0.35(3)  & 0.017(1)  \\
L483                      & +5.46(1) & 0.45(2)  & 0.077(2)    & +5.44(1) & 0.68(2)  & 0.084(2)$^a$    & +5.42(2) & 0.46(2)  & 0.047(2)   & +5.50(3) & 0.58(5)  & 0.021(2)  \\
L1495B                  & +7.78(1) & 0.42(3)  & 0.021(1)    & +7.64(1) & 0.46(2)  & 0.036(1)$^a$    & +7.76(2) & 0.33(4)  & 0.010(1)   & +7.74(3) & 0.31(6)  & 0.005(1)  \\
L1521F                  & +6.61(1) & 0.43(2)  & 0.053(1)    & +6.50(1) & 0.68(2)  & 0.066(1)$^a$    & +6.58(1) & 0.41(2)  & 0.032(1)   & +6.63(2) & 0.33(5)  & 0.007(1)  \\
Serpens South 1a & +7.65(2) & 0.66(4)  & 0.038(2)    & +7.58(1) & 0.63(1)  & 0.149(2)$^a$    & +7.62(2) & 0.68(4)  & 0.030(1)   & +7.50(8) & 1.15(17) & 0.015(2) \\
L1389                    & $-$4.53(1) & 0.45(3)  & 0.060(3)   & -- & -- & -- & -- & -- & -- & -- & -- & -- \\
L1172                    & +2.91(6) & 0.61(11)  & 0.019(3)     & -- & -- & -- & -- & -- & -- & -- & -- & -- \\
L1251A                  & $-$3.79(6) & 0.47(11)  & 0.018(4) & -- & -- & -- & -- & -- & -- & -- & -- & -- \\
L1512                    & +7.26(2) & 0.24(7)  & 0.014(2)      & -- & -- & -- & -- & -- & -- & -- & -- & -- \\
\hline
\end{tabular}
\tablenoteb{\\
Numbers in parentheses are 1$\sigma$ uncertainties in units of the last digits. Line parameters obtained by gaussian fits to the line profiles. $^a$ Uncertain intensity due to severe overlap with the $J=15-14$ transition of C$_3$S at 86708.374 MHz.
}
\end{table*}
}

\onltab{6}{
\begin{table*}
\caption{Observed line parameters of CH$_2$CHCH$_3$} \label{table:ch2chch3}
\centering
\tiny
\begin{tabular}{l@{\hspace{0.1cm}}r | @{\hspace{0.1cm}}c@{\hspace{0.25cm}}c@{\hspace{0.25cm}}c | @{\hspace{0.1cm}}c@{\hspace{0.25cm}}c@{\hspace{0.25cm}}c | @{\hspace{0.1cm}}c@{\hspace{0.25cm}}c@{\hspace{0.25cm}}c@{\hspace{0.1cm}} | @{\hspace{0.1cm}}c@{\hspace{0.25cm}}c@{\hspace{0.25cm}}c}
\hline \hline
\multicolumn{2}{c}{}                 & \multicolumn{3}{c}{\small{Lupus-1A}}           & \multicolumn{3}{c}{\small{L1495B}}        & \multicolumn{3}{c}{\small{L1521F}}                 & \multicolumn{3}{c}{\small{Serpens South 1a}}     \\
\hline
Transition          & Frequency & $V_{\rm LSR}$ & $\Delta v$ & $\int T_A^* dv$ & $V_{\rm LSR}$ & $\Delta v$ & $\int T_A^* dv$ & $V_{\rm LSR}$ & $\Delta v$ & $\int T_A^* dv$ & $V_{\rm LSR}$ & $\Delta v$ & $\int T_A^* dv$ \\
  & (MHz) & (km/s) & (km/s) & (K km/s) & (km/s) & (km/s) & (K km/s) & (km/s) & (km/s) & (K km/s) & (km/s) & (km/s) & (K km/s) \\
\hline
$5_{1,5} - 4_{1,4}$ A & 84151.670   & +5.00(4) & 0.33(11) & 0.005(1)         & +7.55(2) & 0.25(8) & 0.006(1)    & +6.33(4) & 0.34(9) & 0.006(1)             &    --             &    --         &    --               \\
$5_{0,5} - 4_{0,4}$ A & 86651.566   & +5.05(5) & 0.36(9)  & 0.004(1)$^a$ & +7.50(2) & 0.28(4)  & 0.009(1)   & +6.28(2) & 0.17(8)  & 0.005(1)            & +7.46(6)     & 0.58(12) & 0.011(2)       \\
$5_{2,4} - 4_{2,3}$ A & 87134.576   &     --        &      --       &         --            & +7.64(3) & 0.26(7)  & 0.004(1)   &       --      &    --         &       --                &    --             &    --         &    --                \\
$6_{0,6} - 5_{0,5}$ A & 103689.979 & +4.99(7) & 0.52(16) & 0.006(2)         & +7.55(3) & 0.44(8) & 0.008(1)    & +6.25(4) & 0.44(8) & 0.006(1)             &    --             &     --        &    --               \\
$5_{1,5} - 4_{1,4}$ E & 84151.52$^b$ &    --         &    --         &      --           & +7.67$^b$ & 0.25(7) & 0.006(1) & +6.40$^b$ & 0.44(20) & 0.005(2)$^c$ &    --            &     --        &     --              \\
$5_{0,5} - 4_{0,4}$ E & 86650.873 & +5.03(4) & 0.56(10)  & 0.011(2)          & +7.61(2) & 0.33(6)  & 0.010(1)   & +6.34(3) & 0.28(5)  & 0.005(1)             & +7.57(4)    & 0.68(10) & 0.017(2)        \\
$5_{2,4} - 4_{2,3}$ E & 87137.941   &    --         &   --          &    --                 & +7.70(4) & 0.44(9)  & 0.006(1)   &     --        &   --          &     --                   &    --            &      --       &           --          \\
$6_{0,6} - 5_{0,5}$ E & 103689.117 & +5.19(8) & 0.77(16) & 0.009(2)         & +7.64(2) & 0.27(4) & 0.006(1)    & +6.41(3) & 0.30(5) & 0.006(1)              &    --            &      --       &       --              \\
\hline
\end{tabular}
\tablenoteb{\\
Numbers in parentheses are 1$\sigma$ uncertainties in units of the last digits. Line parameters obtained by gaussian fits to the line profiles. $^a$ Intensity diminished due to overlap with frequency switching artifact. $^b$ Rest frequency of transition $5_{1,5} - 4_{1,4}$ of E species constrained to 84151.52 $\pm$ 0.03 MHz by fixing $V_{\rm LSR}$. $^c$ Line blended. %$^d$ Marginal detection.
}
\end{table*}
}

\end{document}